\documentclass{edp-conf}
\usepackage{graphicx,epsfig}
\begin{document}

\TitreGlobal{SF2A 2004}

\title{Precise measurement of CMB polarisation from Dome-C: the BRAIN and 
CLOVER experiments}
\author{Piat, M.}\address{APC, Coll\`ege de France, 11 place Marcelin 
Berthelot, 75231 PARIS}

\author{Rosset, C.$^1$ on behalf of the BRAIN-CLOVER collaboration}
\address{APC Univ. Paris 7, LISIF Univ. Paris 6, LERMA Observatoire de Paris, CESR Toulouse, IAS Orsay, 
Universit\`a di Roma La Sapienza, 
Universit\`a di Milano Bicocca, University of Wales 
Cardiff, Cambridge University }
%
\runningtitle{BRAIN-CLOVER }
%
\index{Piat, M.}
\index{Rosset, C.}

\maketitle
\begin{abstract}
    The characterisation of CMB polarisation is one of the next
    challenges in observational cosmology.  This is especially true
    for the so-called B-modes that are at least 3 orders of magnitude
    lower than CMB temperature fluctuations.  A precise measurement of
    the angular power spectrum of these B-modes will give important
    constraints on inflation parameters. We
    describe two complementary experiments, BRAIN and CLOVER,
    dedicated to CMB polarisation measurement.  These experiments are
    proposed to be installed in Dome-C, Antarctica, to take advantage
    of the extreme dryness of the atmosphere and to allow long
    integration time.
\end{abstract}
%
\section{Introduction}
What is now
known as the standard cosmological model (or concordance model) 
predicts the existence of primordial
gravitational waves, generated during an explosive period of expansion
known as inflation.  This prediction remains yet to be tested and is
the motivation for our proposed experiments at Dome C: to measure the
so-called B-modes of the CMB polarization which are a signature of
primordial gravity waves.  The detection of gravity waves from
inflation would not only be a major discovery for cosmology, but also
for all of fundamental Physics.  It would satisfy one of the primary
objectives of modern Physics to detect these waves predicted by
Einstein's General Relativity.  It would furthermore be an
indication of the quantum nature of gravity, for the excitation
mechanism of inflation is inherently semi-classical (usually reserved 
otherwise for quantum fields).  
Finally, these B-modes would offer
direct access to physics at energies inaccessible in laboratories
and probably related to grand unification.  For example, the amplitude
of the B-mode would immediately give us the energy scale of inflation
and hence the characteristic energy scale of unification physics 
(Knox, 2002).

\section{Observing the CMB polarisation}
The exciting physics
targeted by the B-modes 
requires improvements of up to two orders of magnitude in
sensitivity with respect to the future ESA Planck mission (2007) that 
will map accurately intensity fluctuations and E polarisation of the CMB 
(Bouchet et al. 2003).  
This leads to the need for a CMB polarization
measurement at large angular scale limited only by astrophysical
limits.  Obtaining such sensitivity represents an unequaled
experimental challenge.  Bolometers deliver the best detector
performance, and those being built for the Planck mission will
operate at the photon noise limit. The only way to improve
sensitivity is by increasing the number of detectors. 
It is not, however, only a question of pure detector sensitivity: it is
equally necessary to obtain exquisite control of all systematic errors 
(instrumental parasitic effects) to an unparalleled level. The 
final way to achieve that is by comparing results from independent 
experiments with orthogonal techniques as already happened for CMB 
anisotropy (Spergel et al. 2003).\\ 
Since the measurement is so difficult, independent orthogonal 
experiments are needed together with the very best astronomical site 
on Earth which is 
certainly Dome-C in Antarctica 
(proceedings of "Dome C
Astronomy/Astrophysics Meeting ", 2004).
This site has huge advantages over other
astronomical site, especially for CMB observations: high atmospheric
transmission, low w
ater vapor, stability of the atmosphere (polar
vortex), sun always low in elevation.  Moreover, these exceptional
conditions are available very frequently allowing for very long integration
time.  Note that a variety of experiments have already been or will
soon be operated from Antarctica: Boomerang, ACBAR, DASI, QuaD, 
BICEP.\\
We propose such strategy with the BRAIN/CLOVER program consisting in 
two complementary experiments with different instrumental approach.

\section{The BRAIN experiment: a new instrumental concept} 
\begin{figure}[htb]
    \vspace{-10pt}
    \begin{center}\hbox{
	\psfig{file=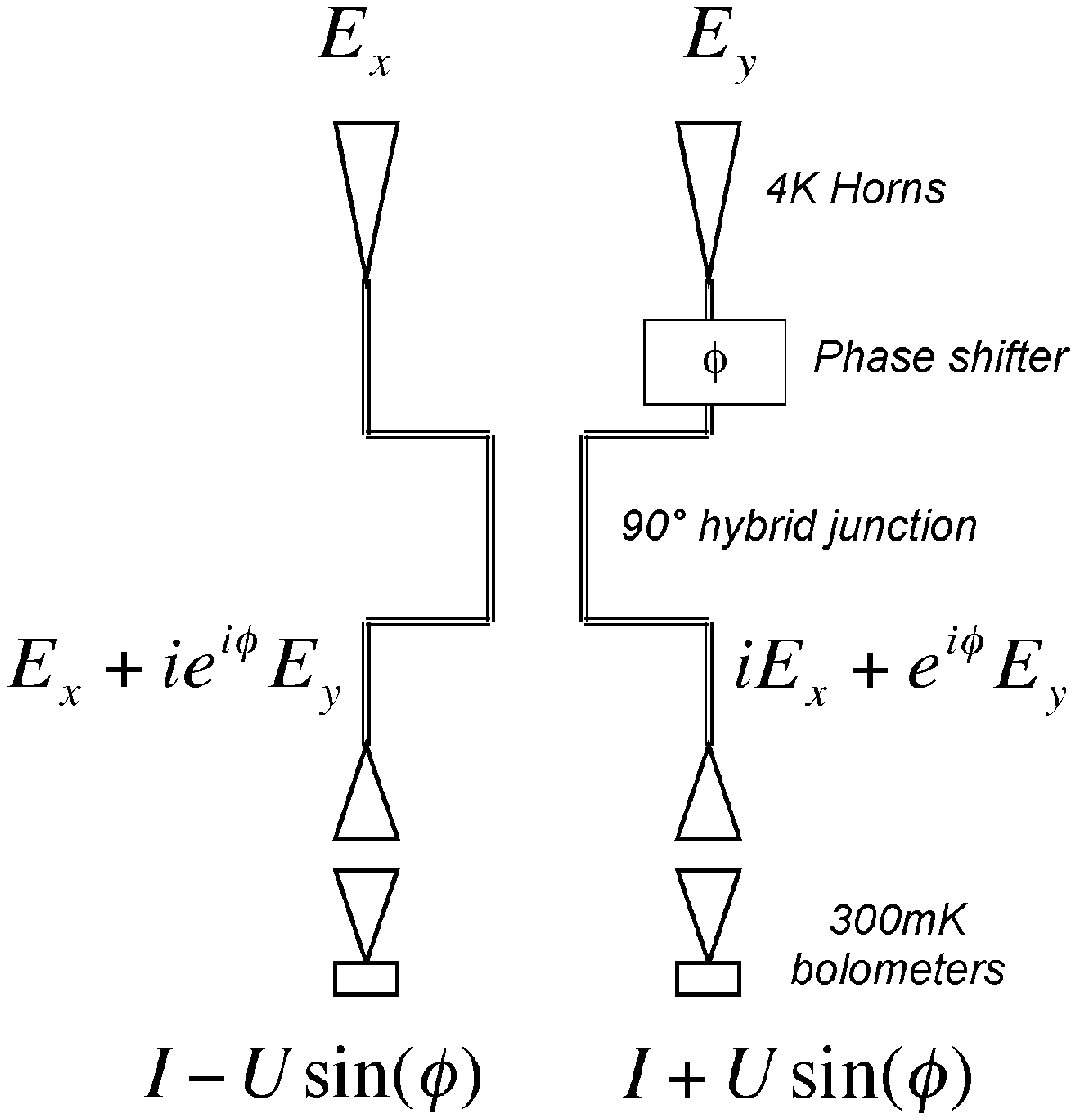, width=55truemm}
	\psfig{file=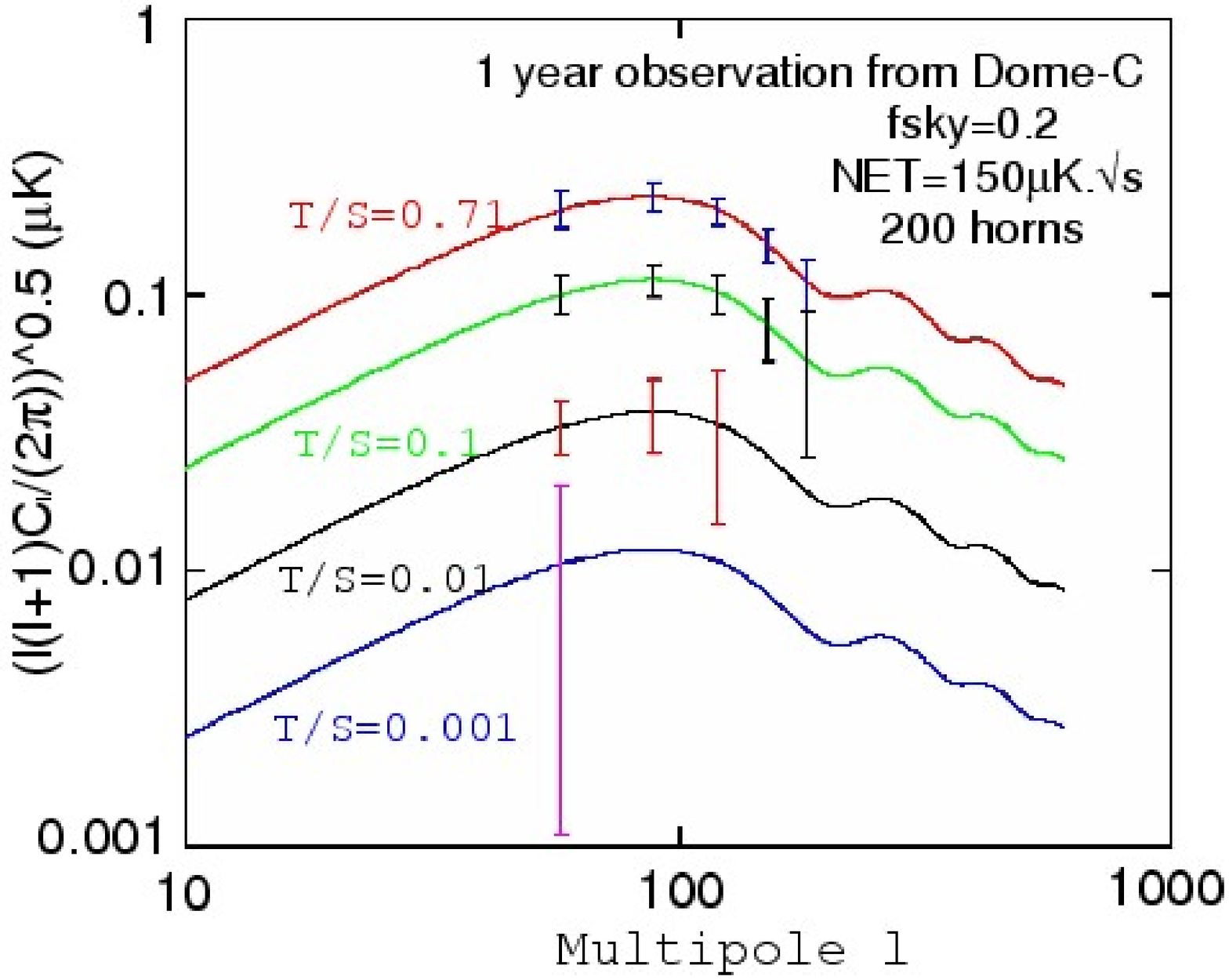, width=65truemm} }
    \end{center} 
    \vspace{-30pt}
    \caption{Left: Principle of polarisation detection with BRAIN (we assumed V=0). 
    Right: Expected errors from BRAIN V2 experiment on the B-mode power 
    spectrum.
    }
    \label{figure_brain}
\end{figure}

Although originally developed to increase angular resolution, aperture
synthesis offers several advantages over direct imaging for
observations of the CMB anisotropies, which include: (i) An 
intrinsically differential measurement, since one
directly obtains the Fourier modes of the sky distribution (this
is particularly advantageous for removing atmospheric signals). 
(ii) The unique sky tracking of the complete set of N antennas of
the array allows the astrophysical signal to be separated from
other sources on spurious signals (e.g., ground pick-up, etcÉ). 
(iii) A direct polarization measurement as output of the
correlator, instead of a determination based on the difference of
two total power measurements.\\
These advantages together with the intrinsic sensitivity of bolometers
are strong motivation for the development of a combined system.  
BRAIN is a prototype of bolometric interferometry which is being 
developed in our international consortium.
It will have 2$\times$2 input horns
or 6 baselines.  One baseline consists of a pair of corrugated
feed-horns, working at 150 GHz, spaced by few wavelengths.  The horns
are directly observing the sky, without telescope, to reduce the sidelobe 
level.
They are cooled to 4K by a 
pulse tube system working continually and
avoiding the use of liquid helium. At the
output of each horn, the 2 polarisations are separated
thanks to an Ortho-Mode Transducer.  They
are then phase-shifted at different frequencies.  All 8
E-M signals from the 4 horns enter a Butler combiner that feeds 8
bolometers cooled to 300mK thanks to a $^3$He fridge. 
A lock-in detection is used to recover the scientific signal contained at 
the beating frequencies of each phase-shifters and 
corresponding to the Fourier transform of one Stokes parameter on the
field of view at a spatial frequency given by the length of the
considered baseline. A simple model of the detection chain is shown 
figure \ref{figure_brain}.\\
It can be shown that such instrument has the same sensitivity of a 
bolometric imager with the same number of detector, with the 
additional immunity to systematic effects. It also allows a measure 
of all 4 Stokes parameters simultaneously with a simpler data 
processing since no map-making process is needed to recover the 
spatial power spectrum. \\
This instrument is proposed to be installed in Dome-C during the 
Antarctic summer 2005-06. It will be followed by a BRAIN version 2 
with 3 bands (90, 150, 220GHz) each with 256 horns to look for B-modes 
(fig. \ref{figure_brain}). This complete instrument could be 
ready for operation in 2008 with a major 
contribution from French and Italian teams.
\section{The CLOVER experiment}
\begin{figure}[htb]
     \begin{center}\hbox{
	\psfig{file=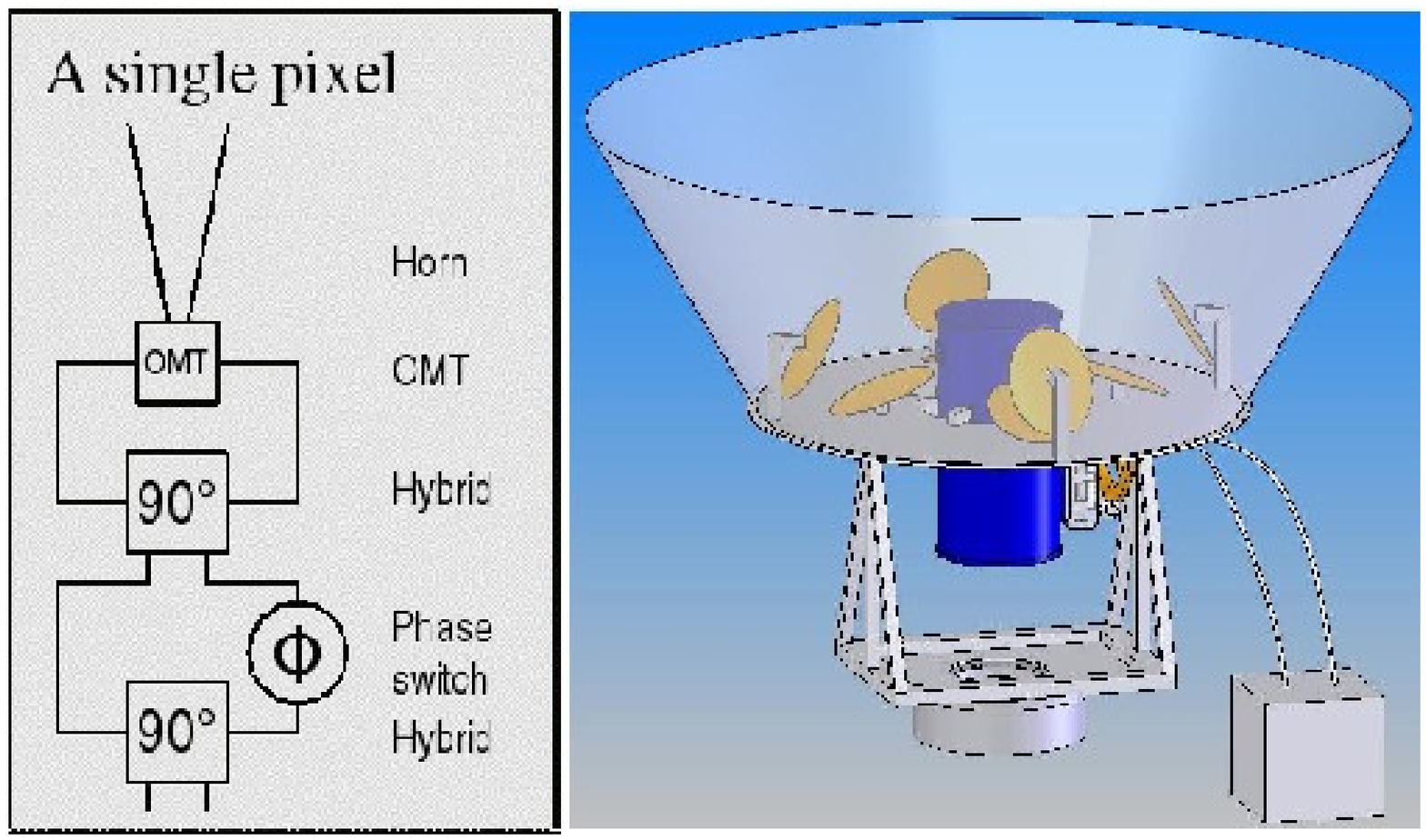, width=72truemm}
	\psfig{file=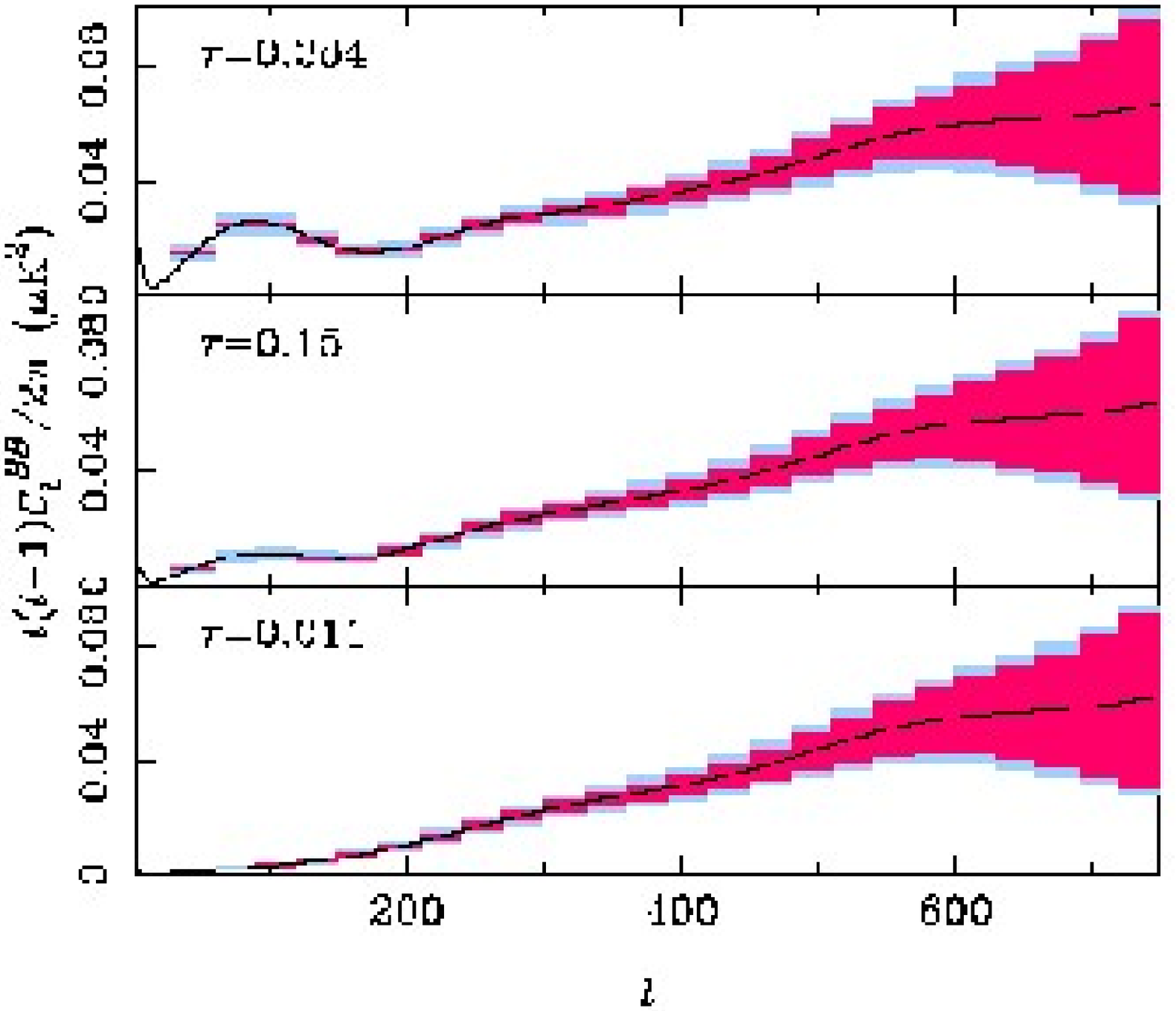, width=48truemm} }
    \end{center} 
    \vspace{-30pt}
    \caption{Left: Principle of the detection chain of CLOVER 
    (pseudo-correlator scheme). Middle: CAD view of one CLOVER 
    channel with its 4 telescopes. Right: 
    Expected errors from CLOVER on the B-mode power spectrum with 
    different tensor-to-scalar ratio r.}
    \label{figure_clover}
\end{figure}

%
CLOVER is a bolometric imager aimed at maping the primordial B-modes
together with the lensed E-modes converted into B-modes.  These lensed
B-modes can lead to complementary cosmological information, such as
neutrino masses.  The CLOVER experiment uses 4 telescopes at 90 deg of
the cryostat that image the same part of the sky on 4 arrays of 64
horns.  Each detection chain is based on a pseudo-correlator scheme (fig. \ref{figure_clover}). 
It uses similar microwave components as BRAIN and allows to obtain for
each pixel a EM signal that is proportional to one of the Stokes
parameter Q or U. The EM signal is furthermore converted into an
electrical current with a finline structure.  Each 4 currents coming
from the same pixel on the sky are added and dissipated in a matched
resistance on a small TES bolometer.  This strategy allows to increase
the signal-to-noise ratio with a relatively small number of detectors. 
The final CLOVER instrument will map the sky in 3 bands (90, 150,
220GHz) each with 4 telescopes and 64 TES. The complete instrument should
be ready for operation in 2008. 
\section{Conclusion}
The BRAIN-CLOVER program is aimed to find the CMB B-modes polarization 
from Dome-C, Antarctica, in the 2008 horizon. It is based on two 
othogonal but complementary experiments that will also give 
important results on polarised foregrounds. This program is a first 
phase to develop a dedicated Post-Planck satellite to be launched 
around 2015-2020. 

\end{document}